\documentclass[aps,pra,twocolumn,superscriptaddress,showpacs,eqsecnum,nofootinbib]{revtex4}
\usepackage{comb,amsmath, latexsym, amssymb, amscd}
\numberwithin{equation}{section}
\usepackage{graphicx}
\usepackage{pictex}

\begin{document}

\title{Optimal cloning of mixed Gaussian states}

\author{M\u{a}d\u{a}lin Gu\c{t}\u{a}}
\affiliation{University of Nijmegen, Toernooiveld 1, Postbus 9010, 6500 GL Nijmegen, The Netherlands} 
\author{Keiji Matsumoto}
\affiliation{National Inatitute of Informatics,
  2-1-2, Hitotsubashi, Chiyoda-ku, Tokyo, 101-8430, Japan}
 \affiliation{Quantum Computation and Information Project, JST,
  Hongo 5-28-3, Bunkyo-ku, Tokyo 113-0033, Japan}

\begin{abstract}
We construct the optimal 1 to 2 cloning transformation for the family of displaced thermal equilibrium states of a harmonic oscillator, with a fixed and known temperature. 
The transformation is Gaussian and it is optimal with respect to the figure of merit based on the joint output state and norm distance. The proof of the result is based on the equivalence between the optimal cloning problem and that of optimal amplification of Gaussian states which is then reduced to an optimization problem for diagonal states of a quantum oscillator. A key concept in finding the optimum is that of stochastic ordering which plays a similar role in the purely classical problem of Gaussian cloning. The result is then extended to the case of $n$ to $m$ cloning of mixed Gaussian states.
\end{abstract}
\maketitle

\section{Introduction}

The no-cloning theorem states that quantum information cannot be copied, i.e. there exists no quantum device whose input is an arbitrarily prepared quantum system and the output consists of two quantum systems whose individual states coincide with that of the input \cite{Dieks,Wootters&Zurek,Yuen,Barnum&Caves&Fuchs&Josza}. 
This and other quantum no-go theorems play an important role in quantum information theory and there exist deep connections with problems in quantum cryptography such as that of eavesdropping \cite{Scarani&Iblisdir&Gisin&Acin}. 
With applications in mind, it is more interesting to derive a quantitative version of the theorem which says how good an approximate cloning machine can do, by providing lower bounds for the error made by any such machine. 

The quality of the approximate clones can be judged either locally, by comparing the state of each individual clone with the input state, or globally by comparing the joint state of the approximate clones with that of independent perfect clones. Note that, because the no-cloning theorem requires that each individual system has the same marginal state as the input, it is the local quality criterion which captures more of its flavor. However if we are interested in the joint state of the output then the global criterion is more useful as it takes into account the correlations between the systems.

Before stating our cloning problem we would like to mention a few important results in this area and we refer to the review  \cite{Scarani&Iblisdir&Gisin&Acin} for a more detailed discussion. The problem of universal cloning for finite dimensional pure 
states was analyzed and solved in \cite{Buzek&Hillery, Gisin&Massar,Bruss&Divincenzo&Ekert&Fuchs,Werner,Keyl&Werner,Buzek&Hillery2,
Cerf, Cerf2,Cerf3}. 
Interestingly, when the figure of merit is the supremum over all input states of the fidelity between the ideal and the approximate clones, it was shown that the same cloning machine is optimal from both the local and the global point of view \cite{Werner,Keyl&Werner}. 

In the case of continuos variables systems the Gaussian states have received a special attention due  to their importance in quantum optics, quantum communication and 
cryptography \cite{Grosshans}. Problems in quantum information such as entanglement measures \cite{Wolf&Gidke&Kruger} and quantum channels \cite{Serafini&Eisert&Wolf} have been partially solved by restricting to the framework of Gaussian states and operations. For coherent states the optimal cloning problem has been investigated in \cite{Lindblad,Cerf&Ipe&Rottenberg, Cerf&Iblisdir} under the restriction of Gaussian transformation. The question whether the optimal cloning map is indeed Gaussian has been answered positively in the case of global figure of merit with fidelity, and 
negatively for the individual figure of merit \cite{Cerf&Kruger&Navez&Werner&Wolf}.

As noticed in \cite{Scarani&Iblisdir&Gisin&Acin}, the area of optimal cloning for mixed states is virtually open partly due to the technical difficulties compared with the pure state case. However we should mention here the phenomenon of ``super-broadcasting'' \cite{D'Ariano&Perinotti,Andersen,D'Ariano&Perinotti&Sacchi} which allows not only perfect  $n$ to $m$ (local) cloning of mixed states but even purification, if $n$ is large enough and the input states are sufficiently mixed. This happens however at the expense of creating big correlations between the individual clones, just as in the case of classical copying.

Quantum cloning shows some similarities to quantum state estimation, for example the pure state case is easier  than the mixed case in both contexts. Recently it has been shown that local cloning for pure states is asymptotically equivalent to estimation 
\cite{Acin}. This paper makes another step in this direction by pointing out that global cloning has a natural statistical interpretation. The statistics literature dedicated to the classical version of this problem \cite{Torgersen} has been an inspiration for this paper and may prove to be useful in future quantum investigations.

The problem which we investigate is that of optimal $1$ to $2$ 
cloning of {\it mixed} Gaussian states using a {\it global} figure of merit. We show that the optimal cloner is Gaussian and is similar to the optimal one for the pure state case. Our figure of merit is based on the {\it norm distance} rather than fidelity, the latter  being more cumbersome to calculate  in the case of mixed states. 
However the result holds as well with other figures of merit such as total variation distance between the distributions obtained by performing quantum homodyne tomography measurements.

In quantum state estimation it has been shown \cite{Guta&Kahn} that the family 
of mixed Gaussian states appears as asymptotic limit of multiple mixed qubit states. 
Based on this result it can be proved \cite{Guta&Matsumoto} that the problem of $n$ to $2n$ global cloning of mixed qubit states is asymptotically equivalent to that of $1$ to 
$2$ cloning of mixed Gaussian states which is addressed in this paper. 


In deriving our result we have transformed the optimal cloning problem into 
an optimal amplifying problem and then used covariance arguments to restrict the 
optimization to the set of mixed number states of the idler of a non-degenerate 
parametric amplifier \cite{Caves,Walls&Milburn}. The argument leading to the conclusion that the optimal state of the idler is the vacuum, is based on the notion of {\it stochastic ordering} which is also used in deriving the solution to the classical problem of 
optimal Gaussian cloning.

In Section \ref{sec.ntom} we extend the solution of the $1$ to $2$ cloning problem to the case of optimal $n$ to $m$ cloning of mixed Gaussian states. The transformation involves three steps: one first concentrates the $n$ modes in one by means of a unitary Fourier transform, then amplifies this mode with a phase-insensitive linear amplifier 
with gain $G= m/n$, and  finally the amplified state is distributed over the $m$ output modes by using another Fourier transform with $m-1$ ancillary modes prepared in a thermal equilibrium state identical to that of the input.

\section{Cloning of mixed Gaussian states}

We consider the problem of optimal cloning for a family of Gaussian states of a quantum oscillator, namely the displaced thermal equilibrium states with a given, known temperature. Let $ a, a^{*}$ be the creation and annihilation operators acting on 
the Hilbert space $\mathcal{H}$ and satisfying the commutation 
relations  $[a, a^{*}]= \mathbf{1}$, and let  
$$
\Phi^{\bf 0}:=(1-s)\sum_{n=0}^{\infty} s^{n} |n\rangle\langle n|,
$$ 
be a thermal 
equilibrium state  where $0<s<1$ is related to the temperature by $s=e^{-\beta}$, 
and $ |n \rangle$ represent the Fock basis vectors of $n$-photons states. 
Let 
$$
\Phi^{\alpha} := D({\alpha}) \Phi^{\bf 0} D({\alpha})^{*},
$$
be the displaced thermal states where 
$$D(\alpha) := \exp (\alpha a^{*} - \bar{\alpha}a),$$ 
and consider the quantum statistical models:
$$
\mathcal{F} := \{ \Phi^{\alpha}   : \alpha\in \mathbb{C} \}, \qquad 
\mathcal{G} := \{ \Phi^{\alpha} \otimes \Phi^{\alpha} : \alpha\in \mathbb{C} \}.
$$
In the next Subsection we will give a statistical interpretation to the optimal figure of 
merit for cloning as a kind of gap (deficiency) between the less informative model 
$\mathcal{F}$ and the more informative one $\mathcal{G}$.

\subsection{Figure of merit} 
The aim of 1 to 2 global cloning is to transform the state $\Phi^{\alpha}$ into  $\Phi^{\alpha}\otimes \Phi^{\alpha}$ without knowing $\alpha$. This is however impossible, 
and this fact has nothing to do with the quantum no-cloning theorem which is about local cloning. In fact the same phenomenon occurs in classical statistics: 
given one Gaussian random variable whose distribution has 
unknown center, it is impossible to produce two {\it independent} variables with the same distribution. In both classical and quantum set-ups, if this was possible one could determine exactly the displacement by first cloning the state to an infinite number of 
independent states and then estimating the displacement using statistical methods.  

Thus we will try to perform an approximate cloning transformation which is optimal with respect to a given figure of merit. We consider a 
{\it global} criterion rather than a local, individual one. 
The classical version of this problem has been previously considered in mathematical statistics \cite{Torgersen} and we will adopt here the same terminology by defining the 
{\it deficiency} of the model $\mathcal{F}$ with respect to the model $\mathcal{G}$ as
$$
\delta_{clon}=\delta(\mathcal{F}, \mathcal{G}) :=\inf_{T} \sup_{\alpha\in \mathbb{C}} \|T(\Phi^{\alpha}) - \Phi^{\alpha} \otimes \Phi^{\alpha}\|_{1}
$$
where the infimum is taken over all possible cloning maps 
$T: \mathcal{S}(\mathcal{H}) \to\mathcal{S}(\mathcal{H})\otimes\mathcal{S}(\mathcal{H}) $ with  $\mathcal{S}(\mathcal{H})$ denoting the space of states (density matrices) on 
$\mathcal{H}$ and $T$ is a completely positive and trace preserving map. 
The norm one of a trace class operator is defined as $\|\tau\|_{1}: =\mathrm{Tr}(|\tau|)$.
We are looking for a map $T=T_{opt}$ satisfying  
$$
F(T_{opt}):=\sup_{\alpha\in \mathbb{C}} \|T_{opt}(\Phi^{\alpha}) - \Phi^{\alpha} \otimes \Phi^{\alpha}\|_{1} = 
\delta_{clon}.
$$
This figure of merit is very natural from the statistical point of view \cite{Guta&Kahn} 
 and can be related with the fidelity through the two sided inequalities \cite{Fuchs&vanddeGraaf}
 $$
\frac{1}{4}\| \rho- \tau\|_{1}^{2} \leq 
2 - 2\mathrm{Tr}(\sqrt{\rho^{1/2} \tau \rho^{1/2}}) \leq \| \rho- \tau\|_{1} . 
$$  
Although the fidelity is a popular figure of merit, it is more difficult to handle in the case of mixed states. Note also that we do not take any average with respect to a prior distribution over the unknown parameter but just consider the cloner which performs best with respect to {\it all} $\alpha$, i.e. we are in a minimax framework as in \cite{Cerf&Kruger&Navez&Werner&Wolf}.

 \subsection{Cloning versus amplifying} \label{subsec.cloningvsamplifying}
 In the classical case the problem of optimal Gaussian cloning is equivalent to that of 
``amplifying'' the location of the center of a Gaussian variable. We will show that this is also the case in the quantum setup by proving a fairly simple lemma allowing us 
to simplify the problem and make the connection with the theory of linear 
amplifiers beautifully exposed in \cite{Caves}.  

Let us start with the classical case and suppose that we 
draw a real number $X$ from the normal distribution $N(u, I)$ with unknown center 
$u\in \mathbb{R}$ and fixed  and known variance $I$. We would like to devise some statistical procedure (for this purpose we may use an additional source of randomness) whose input is $X$ and the output is a pair of independent clones $(Y,Z)\in \mathbb{R}^{2}$, each having distribution  $N(u, I)$. Let us assume for the moment that this can actually be done and note that by performing the invertible transformation 
$$
(Y,Z) \to \left((Y+Z)/\sqrt{2}, \,(Y-Z)/\sqrt{2}\right),
$$
no statistical information is lost and moreover the two newly obtained terms are independent, $(Y+Z) /\sqrt{2}$ has distribution $N(\sqrt{2}u, I)$ while  $(Y-Z)/\sqrt{2}$ does not contain any statistical information about $u$. Conversely, starting from an ``amplified'' version of $X$, that is a variable $\tilde{X}$ with 
distribution $N(\sqrt{2}u, I)$, one can recover the independent clones  by adding an subtracting an independent $N(0,I)$ variable $R$:   
$$
Y= \frac{\tilde{X}+ R}{\sqrt{2}} ,\qquad Z= \frac{\tilde{X}- R}{\sqrt{2}}.
$$
In fact this is nothing else than the classical analogue of a 50-50 beamsplitter 
where $Y,Z$ should be replaced by independent input modes carrying Gaussian states, and $\tilde{X}, R$ can be seen as the output fields. The moral of this is 
that perfect cloning would be equivalent to perfect amplifying if any of them was possible, but in fact the two problems also are equivalent when we content ourselves with finding the optimal solution. We will prove this now in the quantum framework. 
Let $A: \mathcal{S}(\mathcal{H}) \to \mathcal{S}(\mathcal{H})$ be a completely positive, trace preserving channel and 
define the figure of merit for $\sqrt{2}$ amplification
$$
F_{amp}(A) := \sup_{\alpha\in\mathbb{C}} \|A(\Phi^{\alpha}) - \Phi^{\sqrt{2}\alpha} \|_{1},
$$ 
and let $\delta_{amp}:= \inf_{A}F_{amp}(A)$ be the optimal figure of merit.
\begin{lemma}\label{lemma.cloning-amplifier}
If 
$T: \mathcal{S}(\mathcal{H})^{\otimes 2} \to \mathcal{S}(\mathcal{H})$ is an optimal 1 to 2 cloning map then the map
$\rho\mapsto  \mathrm{Tr}_{1} (B \circ T (\rho))$ is an 
optimal $\sqrt{2}$ amplifier, where 
$B: \mathcal{S}(\mathcal{H})^{\otimes 2} \to \mathcal{S}(\mathcal{H})^{\otimes 2}$ is 
the beamsplitter transformation which in the Heisenberg picture is given by the linear transformation 
\begin{displaymath}
B^{*}=
\left(
\begin{array}{cc}
\sqrt{2}      &    -\sqrt{2} \\
\sqrt{2}  &    \sqrt{2}\\
\end{array}\right),
\end{displaymath}
acting on the creation and annihilation operators of the two modes.
Conversely, if $A$ is an optimal amplifier, then the channel
$\rho\mapsto  B^{-1} (A(\rho) \otimes \Phi^{0})$ is optimal for cloning, and in particular  
$\delta_{clon}= \delta_{amp}$.
\end{lemma}

\noindent{\it Proof.} 
Let $A$ be an optimal amplifier, i.e. $F_{amp}(A) = \delta_{amp}$, and $T$ the corresponding cloning map, then
\begin{eqnarray*}
&&
F_{clon}(T) =\\
&&
=\sup_{\alpha\in\mathbb{C}} \|  B^{-1} (A(\phi^{\alpha}) \otimes \Phi^{0})-  
B^{-1} (\Phi^{\sqrt{2}\alpha} \otimes \Phi^{0})\|_{1} \\
&& 
=\sup_{\alpha\in\mathbb{C}} \|A(\phi^{\alpha})- \Phi^{\sqrt{2}\alpha}\|_{1}= F_{amp}(A) = \delta_{amp}.
\end{eqnarray*}
Now let us suppose that there exists another clonig map $T^{\prime}$ with 
$F_{clon}(T^{\prime})< F_{clon}(T)$, then the corresponding amplifier $A^{\prime}$ satisfies
\begin{eqnarray*}
&&
F_{amp}(A^{\prime}) = \sup_{\alpha\in\mathbb{C}} \| \mathrm{Tr}_{1}(B\circ T^{\prime}(\Phi^{\alpha})) -  \mathrm{Tr}_{1}(B (\Phi^{\alpha} \otimes \Phi^{\alpha}))\|_{1} \\
&&
\leq \|T^{\prime}(\Phi^{\alpha}) - \Phi^{\alpha} \otimes \Phi^{\alpha}\|_{1} = 
F_{clon}(T^{\prime}) < F_{clon}(T) = \delta_{amp}.
\end{eqnarray*}
 But this is in contradiction with the definition of the optimal figure of merit for amplification. A similar argument can be applied in the other direction. 
 
 \qed

\subsection{Covariance}\label{sec.covariance}
As in other statistical problems the search for an optimal solution can be simplified if we can restrict the optimization set by means of a covariance argument.


If the cloning map $T$ has the property that if we first displace the input and then apply 
$T$, is equivalent to first applying $T$ and then displacing the outputs by the same amount, then we say that $T$ is (displacement) {\it covariant}:
$$
T(\rho) = D(-\alpha)^{\otimes 2} T(D(\alpha) \rho D(-\alpha)) D(\alpha)^{\otimes 2} := 
T_{\alpha} (\rho),
$$
for all $\alpha\in\mathbb{C}$ and $\rho\in\mathcal{S}(\mathcal{H})$. By convexity of 
the $\|\cdot \|_{1}$ distance we have 
\begin{eqnarray*}
&&
F_{clon}(T) = \sup_{\alpha} \| T(\Phi^{\alpha}) - \Phi^{\alpha}\otimes \Phi^{\alpha} \|_{1} 
\geq 
\\
&&{\bf M}_{\alpha}  \| T_{\alpha} (\Phi^{0}) - \Phi^{0}\otimes \Phi^{0} \|_{1}
\geq \| {\bf M}_{\alpha}T_{\alpha} \Phi^{0} - \Phi^{0} \otimes \Phi^{0}\|_{1} ,
\end{eqnarray*}
where $M_{\alpha}$ is the ``mean with respect to $\alpha$'', the analogue of 
averaging  with respect to an invariant probability measure for the case of compact spaces \cite{Cerf&Kruger&Navez&Werner&Wolf,Torgersen}. Thus the mean ${\bf M}_{\alpha}T_{\alpha}$ is at least as good as the initial channel $T$. There is a technical point here concerning the fact 
that ${\bf M}_{\alpha}T_{\alpha}$  may be singular as it is the case for example if $T$ maps all states in a fixed one then ${\bf M}_{\alpha}T_{\alpha}=0$ which is not trace preserving. A more detailed analysis \cite{Torgersen} shows however that such cases can be excluded and one can restrict attention to proper covariant and trace preserving channels.  

It can be shown \cite{Cerf&Kruger&Navez&Werner&Wolf} that the general form of a covariant cloning map is given in the Heisenberg picture by the linear transformations between the input mode 
$a_{in}$ and the output modes $a_{1}$  and $a_{2}$:
$$
a_{1}= a_{in} + (b_{1}^{*} + b_{2})/\sqrt{2}, \qquad 
a_{2}= a_{in} + (b_{1}^{*} - b_{2})/\sqrt{2},
$$
where $b_{1}$, $b_{2}$ are two additional modes whose joint state determines the action of the cloning map.  

The covariance property can be cast in the amplifier framework as well: a map $A$ is a covariant amplifier if
$$
A(\rho) = D(-\sqrt{2}\alpha) A(D(\alpha) \rho D(-\alpha)) D(\sqrt{2}\alpha),
$$
for all $\rho\in\mathcal{S}(\mathcal{H})$ and $\alpha\in\mathbb{C}$.
As shown in Lemma \ref{lemma.cloning-amplifier} an optimal cloner can be transformed into an optimal amplifier by using a 50-50 beamsplitter to recombine the 2 clones and then keeping one of the outgoing modes. For covariant cloning maps as described above in the Heisenberg picture, this leads to the family of covariant amplifiers (see Appendix) with 
$a_{amp}:=(a_{1}+a_{2})/\sqrt{2}$:
\begin{equation}\label{eq.amplifier}
a_{amp} = \sqrt{2}a_{in} + b^{*},
\end{equation}
where the mode $b_{2}$ has been eliminated and the amplifier depends only on the state of the mode $b:=b_{1}$. Taking this into account, we will analyze the optimality problem in its formulation as optimal amplification. 
We will often use the fact that a particular covariant amplifier $A$ is in one to one correspondence with a state $\tau$ of the mode $b$ as specified by the above linear transformation in the Heisenberg picture, and we emphasize this by writing 
$A=A^{(\tau)}$. 

By using a further covariance argument we will show that the search for optimal amplifier can be restricted to states $\tau$ which are mixtures of number states, i.e. states which are diagonal in the Fock vectors basis. Indeed for any displacement covariant amplifier $A$ we have
\begin{equation}\label{eq.Famp}
F_{amp}(A) = \sup_{\alpha} \|A(\Phi^{\alpha}) - \Phi^{\sqrt{2}\alpha}\|_{1} =  
\|A(\Phi^{0}) - \Phi^{0}\|_{1}.
\end{equation}
Let $U_{b}(\theta)= \exp(i\theta N_{b})$ be the phase transformation with 
$N_{b}=b^{*}b$ the number operator of the mode $b$, and define similar phase transformations for the modes $a_{in}$ and $a_{amp}$. The amplifier $A$ is covariant with respect to phase transformations if 
$$
A(\rho) = U_{amp}(\theta)^{*}A(U_{in}(\theta) \rho U_{in}(\theta)^{*})U_{amp}(\theta)= A_{\theta}(\rho).
$$
It is now easy to check that if $A= A^{(\tau)}$ then $A_{\theta}= A^{(\tau_{\theta})}$ where $\tau_{\theta} = U_{b}(-\theta)\tau U_{b}(-\theta)^{*}$. 
Moreover, from \eqref{eq.Famp} we deduce that $F_{amp}(A) = F_{amp}(A_{\theta})$ because the state $\Phi^{0}$ is invariant under phase transformations and thus 
\begin{eqnarray*}
F_{amp}(A) =
&& \int d\theta \|A_{\theta}(\Phi^{0}) - \Phi^{0}\|_{1}  
\\
\geq&&
\| \bar{A} (\Phi^{0}) - \Phi^{0}\|_{1}= F_{amp}(\bar{A}),
\end{eqnarray*}
where $\bar{A}= \int A_{\theta}d\theta= A^{(\bar{\tau})}$ and $\bar{\tau}$ is the 
phase averaged state $\tau$, i.e. a diagonal density matrix in the number 
operator eigenbasis. The rest of the paper deals with the problem of finding the 
optimal diagonal state for the mode $b$.

 \subsection{Stochastic ordering} 
Let us consider an arbitrary diagonal state 
$\tau := \sum_{n} \tau_{n} |n\rangle\langle n|$ of the mode $b$, and denote 
by $p_{n} = (1-s)s^{n}$ the coefficients of the thermal equilibrium 
 state $\Phi^{0}= \sum_{n} p_{n}|n\rangle\langle n|$ of the $a_{in}$ mode. 
 The state of the mode $a_{amp}$ is itself diagonal and its coefficients can be written 
 as $q^{\tau}_{l}= \sum_{k,r} p_{k} \tau_{r} t^{r}_{lk}$ with  fixed $ t^{r}_{lk}$ coefficients having a complicated combinatorial expression. 
 The optimal  amplifier state $\tau_{opt}$ satisfies
\begin{eqnarray*}
F_{amp} (A^{(\tau_{opt})}) &&= 
\| A^{(\tau_{opt})}(\Phi^{0}) - \Phi^{0} \|_{1} = 
\sum_{l=0}^{\infty} |q^{\tau_{opt}}_{l} -p_{l}| \\ &&
=\inf_{\tau} F_{amp}(A^{(\tau)})=
\inf_{\tau} \sum_{l=0}^{\infty} |q^{\tau}_{l} -p_{l}|. 
\end{eqnarray*}
The problem has been now reduced to the following ``classical'' one: given a 
convex family $\mathcal{D}:=\{ q^{\tau} : \tau\in  \mathcal{S} (\mathcal{H}) \}$ of discrete probability distributions on $\mathbb{N}$ and an additional probability distribution $p$ which does not belong to $\mathcal{D}$, find the closest point in $\mathcal{D}$ with respect to the $\|\cdot\|_{1}$ distance. In general such an optimization problem may 
not have an explicit solution but in our case the notion of stochastic ordering is a key 
tool in finding the optimum. 
\begin{definition}
Let $p=\{p_{l} : l\in \mathbb{N}\}$ and $q= \{q_{l} : l\in \mathbb{N}\}$ be two probability distributions over 
$\mathbb{N}$. 
We say that $p$ is stochastically smaller than $q$ ($p\preceq q$) if 
$$
\sum_{l=0}^{m} p_{l} \geq \sum_{l=0}^{m} q_{l}, \quad \forall m\geq 0.
$$
\end{definition}
The following Lemma is a key technical result which will allow us to identify the optimal amplifier map.
\begin{lemma}\label{lemma.stoch.ordering}
Assume that the mode $a_{in}$ is prepared in the thermal equilibrium state $\Phi^{0}$, and the mode $b$ in an arbitrary diagonal state $\tau$. 
Then the following stochastic ordering holds:  $q^{\omega}\preceq q^{\tau}$ where $q^{\tau}$ is 
the distribution of the mode $a_{amp}$ defined in \eqref{eq.amplifier} and  $\omega= |0\rangle\langle 0|$ is the vacuum state.
\end{lemma}

\noindent{\it Proof.} We will prove the result in two steps. 
First we show that the statement can be reduced to the 
case where the input state is the vacuum rather than a thermal equilibrium state 
$\Phi^{0}$.  Then we prove the lemma for the mode $a_{in}$ in the vacuum state.

In quantum optics the equation \eqref{eq.amplifier} describes a 
 non-degenerate parametric amplifier \cite{Walls&Milburn} whose general input-output transformation has the form
\begin{eqnarray*}
&&
c_{1}(t) = \cosh (\chi t) c_{1} + \sinh(\chi t) c_{2}^{*},\\
&&
c_{2}(t)= \cosh(\chi t) c_{2} + \sinh(\chi t) c_{1}^{*},
\end{eqnarray*}
where $t$ represents the time and $\chi$ is a susceptibility constant. 
If both $c_{1}$ and $c_{2}$ modes are prepared in the vacuum state then each of 
the outputs separately will be in the thermal equilibrium state
$ \cosh (\chi t)^{-2}\sum_{k} \tanh (\chi t)^{2k} |k\rangle \langle k|$. This means that 
we can consider that our input mode $a_{in}$ is one of the {\it outputs} of a parametric amplifier with $\tanh (\chi t)^{2} =s$. Thus
$$
a_{in} = \cosh (\chi t) c_{1} + \sinh (\chi t) c_{2}^{*}.
$$  
which together with \eqref{eq.amplifier} gives
\begin{eqnarray*}
a_{amp} 
&&=
\sqrt{2} \cosh (\chi t) c_{1} + \sqrt{2}\sinh(\chi t) c_{2}^{*} + b^{*} 
\\
&&
=\cosh(\tilde{t}) c_{1} + \sinh (\tilde{t}) ( T c_{2}^{*}  +  R b^{*})\\
&&
=
\cosh(\tilde{t}) c_{1} + \sinh (\tilde{t}) \tilde{c}^{*},
\end{eqnarray*}
where $\cosh(\tilde{t}) =\sqrt{2}\cosh(\chi t)$, $R=\sinh (\tilde{t})^{-1} $ and 
$T= \sqrt{1-R^{2}}$. The right side of the last equation can be interpreted as follows: 
the modes $c_{2}$ and $b$ are combined using a beamsplitter with transmitivity $T$ and one of the emerging beams denoted $\tilde{c}$ is further used together with the mode 
$c_{1}$, as inputs of a parametric amplifier characterized by the coefficient 
$\cosh(\tilde{t})$. By hypothesis we assumed that the mode $b$ is in state $\tau$, and by 
construction the mode $c_{2}$ is in the vacuum, thus  the state of $\tilde{c}$ is given 
by the well known binomial formula \cite{Leonhardt}
$$
\tilde{\tau}=\sum_{k=0}^{\infty} \tau_{k} \sum_{p=0}^{k} \binom{k}{p} T^{2(p-k)} R^{2k}  |p\rangle \langle p| 
= \sum_{p=0}^{\infty} \tilde{\tau}_{p}  |p\rangle \langle p| .
$$
The only property which we need here is that  $\tilde{\tau}$ is the vacuum state if and only if $\tau$ is the vacuum state. In conclusion, by introducing the additional modes $c_{1}$  and $c_{2}$ we have transfered the ``impurity'' of the thermal equilibrium state 
from the mode $a_{in}$ to the mode $\tilde{c}$, and the stochastic ordering statement can be now reformulated in our original notations as follows: the mode $a_{in}$ is prepared in the vacuum, and the mode $b$ is prepared in a state $\tilde{\tau}$ which is equal to the vacuum if and only if $\tau$ is the vacuum. In addition, the relation \eqref{eq.amplifier} should be replaced by 
$$
a_{amp} = \cosh (\tilde{t}) a_{in} + \sinh{\tilde{t}} b^{*}.
$$
Under the assumption that $a_{in}$ is in the vacuum, we proceed with the second step of the proof. Because stochastic ordering is preserved by taking convex combinations,  we may assume without loss of generality that $\tau=|k\rangle\langle k|$ for $k>0$. The following formula \cite{Walls&Milburn} gives 
a computable expression of the output two-modes vector state of the amplifier  
\begin{eqnarray*}
\psi &&= 
e^{\Gamma a_{in}^{*} b^{*}} e^{-g (a_{in}^{*}a_{in}+ b^{*}b+{\bf 1})}
e^{-\Gamma a_{in}b} |0,k\rangle \\
&&=
e^{-g(k+1)} \sum_{l=0}^{\infty} \Gamma^{l} \binom{l+k}{k}^{1/2} |l,l+k\rangle,
\end{eqnarray*}
where $\Gamma = \tanh(\tilde{t})$ and $e^{g}=\cosh(\tilde{t})$. By tracing over the
 mode $b$ we obtain the desired state of $a_{amp}$
 $$
\sum_{l=0}^{\infty} q^{k}_{l} |l \rangle \langle l|= e^{-2g(k+1)}  
\sum_{l=0}^{\infty} \Gamma^{2l} \binom{l+k}{k} |l \rangle 
 \langle l|  $$.
 The relation $q^{\omega}\preceq q^{\tau}$ reduces to showing that
 $$
 \sum_{l=0}^{m}  q^{0}_{l} \geq \sum_{l=0}^{m}  q^{k}_{l},
 $$
 for all $m$.
With the notation $\gamma=\Gamma^{2}$ we get
\begin{eqnarray*}
&&
\sum_{l=0}^{m} q^{k}_{l}=
(1-\gamma)^{k+1} \sum_{l=0}^{m}\gamma^{l} \binom{l+k}{k}\\
&&
=\frac{(1-\gamma)^{k+1}}{k!} \left(\frac{1-\gamma^{k+m+1}}{1-\gamma} \right)^{(k)}\\
&&
=1- \gamma^{m+1} \sum_{r=0}^{k} (1-\gamma)^{r} \gamma^{k-r} \binom{k+m+1}{r}\leq\\
&&
1-\gamma^{m+1} \sum_{r=0}^{k} (1-\gamma)^{r} \gamma^{k-r} \binom{k}{r}=1-\gamma^{m+1} = \sum_{l=0}^{m} q^{0}_{l}.
\end{eqnarray*}

\qed

\begin{lemma}\label{lemma.distance}
We have
$$
\| p- q^{\omega}\|_1 =2 \sup_{m\in \mathbb{N}} \sum_{l=0}^{m} (p_{l} -q_{l})
=2\left(\frac{1+s}{2}\right)^{m_{0}+1} - 2s^{m_{0}+1},
$$
where $m_{0}$ is the integer part of $ \log 2/(\log(1+s)-\log(2s))$.
\end{lemma}

\noindent{\it Proof.} Because both distributions are geometric, there exists an integer 
$m_{0}$ such that $p_{l} \geq q_{l}$ for $m\leq m_{0}$ and $p_{l}< q_{l}$ for 
$m>m_{0}$, and this proves the first equality. From the proof of Lemma \ref{lemma.stoch.ordering} we can compute 
$q^{\omega}_{l}= (1-\tilde{s})\tilde{s}^{l}$ where $\tilde{s}= (1+s)/2$ and thus the integer $m_{0}$ is given by the integer part of the 
$ \log 2/(\log(1+s)-\log(2s))$. In conclusion
$$
\| p- q^{\omega}\|_1 = 2\left(\frac{1+s}{2}\right)^{m_{0}+1} - 2s^{m_{0}+1}. 
$$

\qed

We arrive now to the main result of the paper. We will show that amplifier $A^{\tau}$ whose output is closest to the desired state state, is that corresponding to 
$\tau= |0\rangle \langle0|$. Intuitively this happens because the ``target'' distribution 
$q$ is geometrically decreasing and the closest to it in the family $q^{\tau}$ is the 
output which is the least  ``spread''. This intuition is cast into mathematics through 
the concept of stochastic ordering and the result of Lemma \ref{lemma.stoch.ordering}.

\begin{theorem}\label{th.main}
The state of the mode $b$ for which the corresponding amplifier map $A$ is 
optimal is $\omega = |0\rangle\langle 0|$. In particular, the optimal amplifying and cloning maps are Gaussian. 
\end{theorem}
\noindent{\it Proof.} Define 
$$
m_{a} := \max (m: \sum_{l=0}^{m} p_{l}\leq a) 
$$
and  
$
\mathcal{D}(a, \tau) = \{ D\subset \mathbb{N}: \sum_{l\in D}\tau_{l}\leq a\},
$
for all $a\geq 0$. Note that by Lemma \ref{lemma.stoch.ordering} we have 
$
\sum_{l=0}^{m_{a}} q^{\tau} \leq a$ for all $\tau$, and thus $\{0,1,\dots ,m_{a}\} \in \mathcal{D}(a, \tau)$. Using the relation 
$\|p-q\|_{1} = 2\sup_{D} \sum_{l\in D}(p_{l}- q^{\tau}_{l}) $ we obtain the chain of inequalities
\begin{eqnarray*}
&&
\|p- q^{\tau}\|_{1} = 2 \sup_{a\geq 0}\, \sup_{D \in \mathcal{D}(a,\tau)}\, \sum_{l\in D} 
(p_{l}- q^{\tau}_{l})\\
&& 
\geq 
2\sup_{a\geq 0} \sum_{l=0}^{m_{a}} (p_{l}- q^{\tau}_{l}) 
\geq
2\sup_{a\geq 0} \sum_{l=0}^{m_{a}} (p_{l} - q^{\omega}_{l})\\
&&
=2\sup_{m\geq 0} \sum_{l=0}^{m} (p_{l}-q^{\omega}_{l})= \|p-q^{\omega}\|_{1}.
\end{eqnarray*}
The first equality follows directly form the definition of $\mathcal{D}(a,\tau)$. 
The following inequality restricts the supremum over all 
$D\in\mathcal{D}(a,\tau)$ to one element $\{0,1,\dots, m_{a}\}$. In the second 
inequality we replace the distribution $q^{\tau}$ by $q^{\omega}$ using the stochastic ordering proved in Lemma  \ref{lemma.stoch.ordering}. In the following equality we use the fact that both distributions $p$ and $q^{\omega}$ are geometric (see also Lemma \ref{lemma.distance}).

As discussed in Section \ref{sec.covariance}, we can restrict to covariant 
amplifiers and the figure of merit in this case is simply 
$\| A^{\tau}(\Phi^{0}) -\Phi^{0}\|_{1}= \|q^{\tau} - p\|_{1}$, thus the optimal amplifier is 
$A^{\omega}$. Moreover,by the equivalence between optimal cloning and optimal 
amplification we also obtain the optimal cloning map (see Lemma \ref{lemma.cloning-amplifier}).

\qed

\begin{corollary}
The optimal 1 to 2 cloning figure of merit is
$$
\delta_{clon} = 2\left(\frac{1+s}{2}\right)^{m_{0}+1} - 2s^{m_{0}+1},
$$
with $m_{0}$ as in Lemma \ref{lemma.distance}.
\end{corollary}

\noindent{\it Proof.} This follows from Theorem \ref{th.main} and Lemma \ref{lemma.distance}. 

\qed
 
 \subsection{Comparison with the classical case}
The derivation of our result on optimal quantum cloning is inspired by a similar 
one in the classical domain \cite{Torgersen}. In this subsection we comment on
 the optimal figures of merit in the two cases as function of the parameter $s=e^{-\beta}$. 

It is well known that an arbitrary state $\rho$ of a quantum harmonic oscillator has an alternative representation as a function $W_{\rho}:\mathbb{R}^{2}\to \mathbb{R}$ called the Wigner function. In the case of the family of displaced thermal equilibrium states  the Wigner function is a two dimensional Gaussian \cite{Leonhardt}
$$
W_{q_{0},p_{0}}(p,q) = \frac{1}{\pi} \tanh (\beta/2) e^{-((q-q_{0})^{2}+(p-p_{0})^{2})
\tanh (\beta/2)}
$$
with variance 
\begin{equation*}\label{eq.V1}
V_{s} = \frac{1}{2\tanh (\beta/2)}= \frac{1+s}{2(1-s)},
\end{equation*}
and 
$q_{0}+ip_{0}= \sqrt{2}\alpha$. 
We have shown that the best quantum amplifier produces a Gaussian state with 
$\tilde{s}= (s+1)/2$ or in terms of the variance
\begin{equation*}\label{eq.V2}
V_{\tilde{s}}=  \frac{3+s}{2(1-s)}
\end{equation*}
which implies that for any $s\in [0,1)$  we have the relation 
\begin{equation}\label{eq.Var.ineq}
V_{\tilde{s}} = 2 V_{s} +\frac{1}{2},
\end{equation}
which indicates the least noisy amplification according to in the fundamental theorem for phase-insensitive amplifiers \cite{Caves}.

Let us consider now the classical problem of Gaussian cloning as discussed in the beginning of Subsection \ref{subsec.cloningvsamplifying}: given a Gaussian random variable $X\in\mathbb{R}^{2}$ with distribution $N(u, I)$, we want to produce a pair $(Y,Z)$ of independent clones of $X$. 
By using the equivalence between the cloning and the amplification problems,
 the task is equivalent to that of producing a variable $\tilde{X}$ with distribution 
$N(\sqrt{2}u, I)$, and the optimal solution to this problem \cite{Torgersen} is simply to take $\tilde{X}= \sqrt{2}X$! We note that in the classical case the variance of the 
output is {\it always} equal to the double of the variance of the input, while in the quantum case the output ``noise'' is always higher due to the unitarity conditions imposed by quantum mechanics \cite{Caves}, 
and we recuperate the factor 2 in the 
high temperature limit \eqref{eq.Var.ineq}.

\begin{figure}[htbp!]
\begin{center}
\includegraphics[width=8cm]{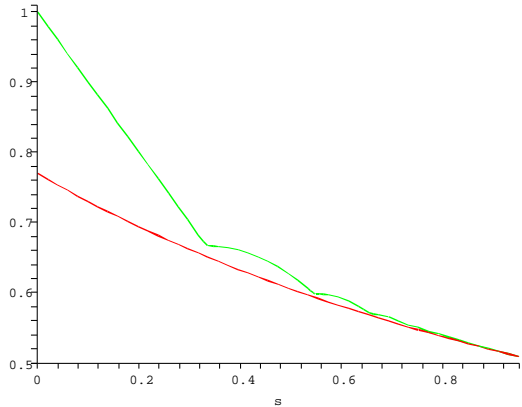}
\caption{Figures of merit for quantum and classical optimal cloning as function of 
$s=e^{-\beta}$.} 
\label{fig.cloning}
\end{center}
\end{figure}

In the classical case one can deduce by a simple scaling argument  that the classical figure of merit does not depend on the variance of the Gaussian but only on the amplifying factor, and in our case it takes the value $1/2$. As expected, the quantum figure of merit is larger than the classical one to which it converges in the limit of high temperature, $s\to 1$. The upper line in Figure \ref{fig.cloning} represents the optimal figure of merit $\delta_{clon}=\delta_{amp}$ as function of $s$. An interesting feature 
of this function is that it appears to have discontinuities in the first derivative precisely at the values of $s$ for which the ``crossing point '' $m_{0}$ makes a jump 
(see Lemma \ref{lemma.distance}). 

For comparison we have also plotted the norm one distance between the 
corresponding Gaussian Wigner functions which does not seem to show any roughness.

\section{Optimal $n$ to $m$ cloning of mixed Gaussian states}\label{sec.ntom}

The results which we obtained for optimal $1$ to $2$ cloning can be easily extended 
to the case of optimal cloning of $n$ to $m$ cloning of mixed Gaussian states. 
The idea is to first ``concentrate'' the state $(\Phi^{\alpha})^{\otimes n}$ of the input into a single mode by means of a unitary transformation followed by discarding the uninteresting $n-1$ modes. Then, one 
amplifies the obtained state by a factor $\sqrt{m/n}$ (gain factor $G=m/n$) and distributes it using another unitary transformation applied on the amplified mode together with $m-1$ additional ancillary modes prepared in state $\Phi^{0}$. The unitary transformations are the Fourier transforms  \cite{D'Ariano&Perinotti&Sacchi}: 
\begin{eqnarray}
&&
 a_{k}  \mapsto U_{n} a_{k} U_{n}^{*}:=\frac{1}{\sqrt{n}} \sum_{l=0}^{n-1} e^{\frac{2\pi ikl}{n}} a_{l},\\
&&
b_{k} \mapsto U_{m} b_{k} U_{m}^{*}:=\frac{1}{\sqrt{m}} \sum_{l=0}^{m-1} e^{\frac{2\pi ikl}{m}} b_{l},
\end{eqnarray}
where $a_{k}$ are the input modes and $U_{m} b_{k} U_{m}^{*}$ are the output 
modes. The amplifying part is described by the covariant map
\begin{equation}\label{eq.amplifier.m}
a_{amp} = \sqrt{\frac{m}{n}} a_{in} + \sqrt{\frac{m-n}{n}} b^{*} 
\end{equation}
where $b$ is an additional mode prepared in a diagonal state $\tau$ as in the $1$ to 
$2$ case, and 
$$
a_{in} := U_{n} a_{0} U_{n}^{*}= \frac{1}{\sqrt{n}} \sum_{l=0}^{n-1} a_{l}.
$$ 
After amplification the second unitary transformation is performed on 
$b_{0}:=a_{amp}$ and the ancillary modes  $b_{1},\dots, b_{m-1}$ prepared in the 
state $\Phi^{0}$.

An obvious extension of Lemma \ref{lemma.cloning-amplifier} holds in this case as well, showing the equivalence of optimal cloning and optimal amplification.  Similarly, 
Lemma \ref{lemma.stoch.ordering} and Theorem \ref{th.main} hold in general for any amplifying factor $k>1$ and we arrive to the conclusion that the optimal amplifier is given by the transformation described in equation \eqref{eq.amplifier.m} with the idler mode $b$ in the vacuum state.

An interesting fact is that our transformations are similar to those of optimal $n$ to $m$
broadcasting \cite{D'Ariano&Perinotti&Sacchi} with the exception that in the last step of the procedure different ancillary states are used: the optimal state for broadcasting is the vacuum while for global cloning it is same thermal equilibrium state $\Phi^{0}$ which characterizes the family.

\section{Concluding remarks}

We have constructed an optimal 1 to 2 cloning map for the family of displaced thermal equilibrium states of a fixed, known temperature. We have considered a global 
figure of merit based on the supremum over all displacements of the norm distance between the joint state of the approximate clones and that of the ideal ones. 
The optimal cloner is Gaussian and is similar with the optimal cloner for coherent 
states with global figure of merit and consists of two operations. The amplification step uses a non-degenerate linear amplifier with idler prepared in the vacuum state. 
The cloning step uses a beamsplitter and another ancillary mode in thermal equilibrium state with the same temperature as the input. 

Computations which have not been included here indicate that the optimal cloning map remains unchanged under global figures of merit using different ``distances'' between states. 

The local version of the optimal cloning problem would probably lead to a non-Gaussian optimum as it is the case with coherent states \cite{Cerf&Kruger&Navez&Werner&Wolf}. 

The equivalence between cloning and amplifying can be extended to an arbitrary 
number $n$ of input states and number of clones $m$, as well as the proof of the optimal amplifier. In the case $n>1$, the first step is the concentration into one mode by means of a unitary Fourier transform, followed by amplification with gain factor 
$G=m/n$, and distribution into $m$ output modes using another Fourier transform.

Some other generalizations of the Gaussian cloning problem may be considered for future investigations, such as an arbitrary number of modes with larger families of Gaussian states. For example in the case of a family of thermal equilibrium states with unknown temperature, one may need to perform an additional estimation of the thermal states in the last step of the cloning which requires ancillary modes prepared in the 
equilibrium state $\Phi^{0}$.

Finally, the key ingredient in our proof was the notion of stochastic ordering which is worth investigating more closely in the context of quantum statistics.

\vspace{1cm}

\noindent{\it Acknowledgments.} We thank Richard Gill and Jonas Kahn for discussions 
and sugesstions. M\u{a}d\u{a}lin Gu\c{t}\u{a} acknowledges the financial support received from the  Netherlands Organisation for Scientific Research (NWO).


\appendix*
\section{Displacement covariant amplifiers}

We give here a short proof of the fact that the displacement covariant amplifiers have the form \eqref{eq.amplifier}. Let 
$A:\mathcal{S}(\mathcal{H}) \to \mathcal{S}(\mathcal{H})$ be  a covariant amplifier such that
$$
A(\rho) = D(-\sqrt{2}\alpha) A(D(\alpha) \rho D(-\alpha)) D(\sqrt{2}\alpha).
$$
Then the dual $A^{*}: \mathcal{B}(\mathcal{H})\to\mathcal{B}(\mathcal{H})$ has a similar property, for all $X\in \mathcal{B}(\mathcal{H})$, $\alpha\in\mathbb{C}$ 
$$
A^{*}(X) = D(\alpha) A^{*}\left(D(-\sqrt{2}\alpha)XD(\sqrt{2}\alpha)\right)D(-\alpha). 
$$
By choosing $X= D(\beta)$ and using the Weyl relations $D(\alpha)D(\beta)= 
\exp(i\mathrm{Im}(\bar \alpha\beta)) D(\alpha+\beta)$  we get 
$A^{*}(D(\beta)) = c(\beta)D(\sqrt{2}\beta)$ for some scalar factor 
$c(\beta)$. Now, according to the Theorem 2.3 of \cite{Daemon&Vanheuverzwijn&Verbeure} if $A$ is trace preserving and 
completely positive the constant $c(\beta)$ is of the form 
$c(\alpha) = \rho(D(\bar \alpha))$ where $\rho$ is a state in 
$\mathcal{S}(\mathcal{H})$. Thus
$$
A^{*}( D(\alpha) ) = \rho(D(\bar \alpha)) D(\sqrt{2} \alpha). 
$$
Now, it can be checked that if we start from \eqref{eq.amplifier} with the mode $b$ prepared in state $\rho$ then $A$ describes the channel transformation from the 
input to the output mode.


\end{document}